\documentclass[pre,twocolumn,aps,showpacs,showkeys]{revtex4}
\usepackage{graphicx}

\begin{document}
\title{
Dissipative discrete breathers in rf SQUID metamaterials
}
\author{N. Lazarides$\ ^{1,2}$, G. P. Tsironis$\ ^{1}$
and M. Eleftheriou$\ ^{1,3}$
}
\affiliation{
$\ ^{1}$Department of Physics, University of Crete, 
and Institute of Electronic Structure and Laser,
Foundation for Research and Technology-Hellas,
P. O. Box 2208, 71003 Heraklion, Greece \\
$\ ^{2}$Department of Electrical Engineering,
Technological Educational Institute of Crete,
P. O. Box 140, Stavromenos, 71500, Heraklion, Crete, Greece \\
$\ ^{3}$Department of Music Technology and Acoustics,
Technological Educational Institute of Crete,
E. Daskalaki, Perivolia,
74100 Rethymno, Crete, Greece
}
\date{\today}
\begin{abstract}
The existence and stability of dissipative discrete breathers (DDBs) in 
rf superconducting quantum interference device (SQUID) arrays in 
both one and two dimensions is investigated numerically.
In an rf SQUID array, the nonlinearity which is intrinsic to each SQUID
due to the presence of the Josephson junction (on-site nonlinearity), 
along with the 
weak coupling of each SQUID to its nearest neighbors through
magnetic forces, result in the appearance of discrete breathers. 
We analyze several discrete breather excitations, both in one and 
two dimensions, which are subjected to unavoidable losses.
These losses, however, are counter-balanced by an external flux
source leading to linearly stable discrete breather structures up to 
relatively large coupling parameters.
We show that DDB excitations may locally alter the magnetic response of
array from paramagnetic to diamagnetic or vice versa,
and that they are not destroyed by increasing the dimensionality.
\end{abstract}

\maketitle
\section{Introduction.
}
Discrete breathers (DBs), also known as intrinsic localized modes (ILMs),
are spatially localized, time-periodic, and stable (or at least long-lived)
excitations in spatially extended, periodic, discrete, nonlinear systems
\cite{Flach,Campbell}. 
They can be produced spontaneously in a nonlinear lattice of weakly coupled 
elements as result of fluctuations \cite{Peyrard,Rasmussen}, 
disorded \cite{Rasmussen1}, 
or by purely deterministic mechanisms \cite{Hennig,Hennig1}.
Since their introduction \cite{Sievers}, a large volume of analytical and 
numerical studies have explored the existence and the properties of DBs
in a variety of nonlinear mathematical models of physical systems.
Rigorous mathematical proofs of existence of DBs in both
energy conserved and dissipative lattices have been given \cite{Mackay,Aubry},
and numerical algorithms for their numerically exact construction 
have been designed \cite{Marin,Marin1,Zueco,Tsironis,Bergamin,Panagopoulos}.
They have been observed experimentally in a variety of physical systems,
including solid state mixed-valence transition metal complexes \cite{Swanson},
quasi-one dimensional antiferromagnetic chains \cite{Schwarz},
arrays of Josephson junctions \cite{Trias},
micromechanical oscillators \cite{Sato}, 
optical waveguide systems \cite{Eisenberg},
layered crystal insulator at $300 K$ \cite{Russell},
and proteins \cite{Edler}.

From the perspective of applications to experimental situations
where an excitation is  subjected to dissipation and external driving,
dissipative DBs (DDBs) are more relevant than their  
Hamiltonian (i.e., energy conserved) counterparts.
Clearly, the dynamics of DDBs is governed by power balance,
rather than energy conservation. In that case, quasiperiodic and even 
chaotic DDBs may exist \cite{Martinez1,Maniadis}.
Recently, DDBs have been demonstrated numerically in discrete 
and nonlinear magnetic metamaterials (MMs) in both one and two dimensions
\cite{Lazarides,Eleftheriou}. 
The MMs are artificial, composite, inherently non-magnetic 
materials that exhibit electromagnetic (EM) properties not available in
naturally occuring materials. 
They are typically made of subwavelength resonant elements like, for
example, the split-ring resonator.
When driven by an alternating EM field, 
the MMs exhibit large magnetic response, either positive or negative,
at frequencies ranging from the microwave up to the Terahertz (THz) and 
the optical bands
\cite{Yen,Podolskiy,Soukoulis}. 
Only a few natural materials respond magnetically at those frequencies,
and that response is usually very weak and within a very narrow band.
Thus, the magnetic response of materials at THz and optical frequencies
is particularly important for the implementation of devices such as
compact cavities, tunable mirrors, isolators, and converters.  
The negative response of MMs can be achieved above the resonance 
frequency, resulting in an effectivelly negative value of the magnetic 
permeability $\mu$, the macroscopic parameter characterizing the magnetic
response of a system. In a linear MM, the effective permeability $\mu$
does not depend on the intensity of the propagating EM field.

In contrast to the linear case, the effective parameters of MMs do depend
on the intensity of the propagating EM field. 
Thus, the nonlinearity offers the possibility to achieve dynamic control 
over  the effective parameters of a MM in real time, 
and thus tuning its properties by changing the intensity of that field.
It has been recently suggested that periodic rf SQUID arrays can operate 
as nonlinear MMs in microwaves, 
due to the resonant nature of the SQUID itself and the nonlinearity inherent
in the Josephson element. In that case, the effective $\mu$ of the rf SQUID
array can be tuned by the applied flux \cite{Lazarides1}.
The combined effects of nonlinearity and discreteness (also inherent
in rf SQUID arrays), may lead in the generation of nonlinear excitations
in the form of DDBs.
In the present context of MMs, such highly localized modes may alter locally
the magnetic response of those materials \cite{Eleftheriou}.

In the present work we investigate numerically the existence and stability 
of DDBs
in both one dimensional (1D) and two-dimensional (2D) arrays of 
rf superconducting quantum interference devices (SQUIDs).
In the next section we describe the two-dimensional rf SQUID array model 
which consists a simple realization of 
a planar MM, while in section III we discuss its linear dispersion properties.
In section IV we construct and present
several types of DDBs both in one and two dimensions. 
In section V we shortly discuss the magnetic response of the rf SQUID 
arrays, showing that DDBs can  locally alter the magnetic response
from paramagnetic to diamagnetic (or vice versa).
We finish in section VI with the conclusions.

\begin{figure}[!t]
\includegraphics[angle=0, width=.5\linewidth]{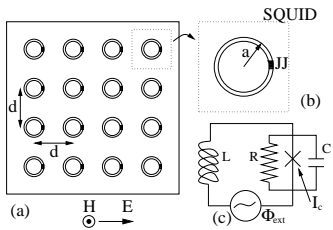}
\caption{
Schematic drawing of the rf SQUID array, along with the equivalent circuit
for a single rf SQUID in external alternating flux $\Phi_{ext}$.  
}
\end{figure}

\section{rf SQUID array model.
}
An rf SQUID,
shown schematically in Fig. 1b, consists of a superconducting ring
interrupted by a Josephson junction (JJ) \cite{Likharev,Barone}.
When driven by an alternating magnetic field, the induced
supercurrents in the ring are determined by the JJ
through the celebrated Josephson relations \cite{Josephson}.
Adopting the resistively and capacitively shunted junction (RCSJ)
model for the JJ \cite{Likharev,Barone}, 
an rf SQUID in an alternating  magnetic field $H_{ext} \equiv H$ 
perpendicular to its plane 
is equivalent to the lumped circuit model shown in Fig. 1c.
That circuit consists of an inductance $L$ in series with an ideal Josephson 
element $I_c$ 
(i.e., for which $I=I_c \sin\phi$, where $I_c$ is the critical current of the JJ
and $\phi$ is the Josephson phase)
shunted by a capacitor $C$ and a resistor $R$, driven by an alternating
flux $\Phi_{ext} (H)$.
The rf SQUID is a nonlinear oscillator which, in an alternating magnetic
field exhibits a resonant magnetic response at a particular frequency
$\omega_0 \simeq 1/ \sqrt{L \, C} = \omega_p/\sqrt{\beta_L}$ 
(for $R\rightarrow \infty)$,
where $\omega_p$ is the plasma frequency of the JJ and 
$\beta_L = 2\pi L I_c / \Phi_0$ is the SQUID parameter
(with $\Phi_0$ being the flux quantum).

Consider a planar rf SQUID array consisting of identical units 
as shown in Fig. 1a, 
arranged in an orthogonal lattice with constants $d_x$ and $d_y$
in the $x$ and $y$ directions, respectively. 
That system is placed in a spatially uniform magnetic field 
$H=H_0 \, \sin(\omega t)$ 
of amplitude $H_{0}$
and frequency $\omega$ ($t$ is the time variable), perpendicular to the SQUID rings.
The field induces a supercurrent $I_{nm}$ in the $nm$th SQUIDs through the flux
$\Phi_{ext} = \Phi_{e0} \, \sin(\omega t)$ threading the SQUID loop 
($\Phi_{e0}=\mu_0 S H_{0} \omega$, where $\mu_0$ is the permeability of the vacuum 
and $S$ the loop area of the SQUID).
The supercurrent $I_{nm}$ produces a magnetic field which couples that SQUID with
all the others due to magnetic dipole-dipole interactions through their 
mutual inductances.
The behavior of that magneto-inductivelly coupled SQUID array is expected 
to deviate 
significantly from directly coupled SQUID arrays like, 
e.g. those studied in Ref. \cite{Chen1}.
Assuming ring-shaped rf SQUIDs of radius $a$ (so that $S=\pi a^2$),
and using nearest-neighbor coupling between them, the flux $\Phi_{nm}$
trapped in the $(n,m)-$th SQUID ring is given by
\begin{eqnarray}
  \label{1}
    \Phi_{nm} = \Phi_{ext} + L \, [ I_{nm} &+&\lambda_x ( I_{n-1,m} + I_{n+1,m} )
     \nonumber \\
     &+&\lambda_y ( I_{n,m-1} + I_{n,m+1} ) ] ,
\end{eqnarray}
where $\lambda_{x,y} \equiv M_{x,y} / L$ 
are the coupling constants between any two 
neighboring SQUIDs in the $x$ and $y$ directions, coupled through mutual
inductances $M_x$ and $M_y$, respectively.
Both values of the $M_x$ and $M_y$ are negative due to the fact that the
magnetic field generated by one SQUID crosses the neighboring SQUID
in the opposite direction.
The supercurrent $I_{nm}$ in the $(n,m)-$th SQUID ring is given, 
within the RCSJ model, by
\begin{equation}
  \label{2}
    -I_{nm} = C\frac{d^2 \Phi_{nm}}{dt^2} +\frac{1}{R} \frac{d \Phi_{nm}}{dt} 
    + I_c\, \sin\left( 2\pi\frac{\Phi_{nm}}{\Phi_0} \right) .
\end{equation}
Due to the planar array geometry and for sufficiently large separations
$d_x$ and $d_y$ in the $x$ and $y$ directions, we may assume that 
$\lambda_x, \lambda_y << 1$, and the nearest neighbor approximation holds.
For the same reasons, we may neglect in the dynamical equations governing
the fluxes in the SQUIDs all those terms of higher order, i.e., terms of the 
form $\lambda_x \lambda_y$, $\lambda_y^2$, $\lambda_x^2$, etc. 

Solving Eq. (\ref{1}) for the current $I_{nm}$  we get 
\begin{eqnarray}
  \label{3}
  I_{nm}= \frac{ \Phi_{nm} - \Phi_{ext}}{L} -\lambda_x ( I_{n-1,m} + I_{n+1,m} )
     \nonumber \\
   -\lambda_y ( I_{n,m-1} + I_{n,m+1} ) .
\end{eqnarray}
Then we  substitute Eq. (\ref{3}), 
written for the currents $I_{n\pm 1,m}$ and $I_{n,m\pm 1}$, 
back into itself.
Omitting higher order terms in the couplings, we get after rearrangement 
\begin{eqnarray}
  \label{4}
    \Phi_{nm} = \Phi_{ext} + L\, I_{nm}
     +\lambda_x ( \Phi_{n-1,m} + \Phi_{n+1,m} -2\Phi_{ext} ) 
     \nonumber \\
     +\lambda_y ( \Phi_{n,m-1} + \Phi_{n,m+1} -2\Phi_{ext}) . 
\end{eqnarray}
By replacing $I_{nm}$ in the earlier equations from Eq. (\ref{2}) we get
\begin{eqnarray}
  \label{5}
    C\frac{d^2 \Phi_{nm}}{dt^2} +\frac{1}{R} \frac{d \Phi_{nm}}{dt}
    + I_c\, \sin\left( 2\pi\frac{\Phi_{nm}}{\Phi_0} \right) + 
    \nonumber \\
    -\lambda_x ( \Phi_{n-1,m} + \Phi_{n+1,m} )
    -\lambda_y ( \Phi_{n,m-1} + \Phi_{n,m+1} )
    \nonumber \\
    = [1 -2 (\lambda_x +\lambda_y )] \Phi_{ext} .
\end{eqnarray}    
Using the relations $f_{nm} = \Phi_{nm} / \Phi_0$,
$f_{ext} = \Phi_{ext} / \Phi_0$, $\beta= \beta_L / 2\pi \equiv L I_c/\Phi_0$,
$\gamma=L \omega_0 /R$, $\tau=\omega_0 t$, equations (\ref{5}) can be written 
in the normalized form
\begin{eqnarray}
  \label{6}
   \frac{d^2 f_{nm}}{d\tau^2} +\gamma \frac{d f_{nm}}{d\tau} +f_{nm}
   +\beta\, \sin( 2 \pi f_{nm} )
    \nonumber \\
   - \lambda_x ( f_{n-1,m} + f_{n+1,m} ) 
   - \lambda_y ( f_{n,m-1} + f_{n,m+1} )
    \nonumber \\
   = [1-2(\lambda_x +\lambda_y)] f_{ext} .
\end{eqnarray}		
Note that the time derivative of $f_{nm}$ corresponds to the voltage $v_{nm}$
across the JJ of the $(n,m)-$th rf SQUID, i.e.,
\begin{eqnarray}
  \label{6.1}
   v_{nm} = \frac{d f_{n,m}}{d\tau} .
\end{eqnarray}   
The small parameter $\gamma$ actually represents all of the dissipation 
coupled to each rf SQUID, which may also include radiative losses \cite{Kourakis}.
Equations (\ref{5})  can be also obtained from the Hamiltonian
\begin{eqnarray}
  \label{7}
   H &=& e^{-t/\tau_C} \,  \sum_{n,m} \frac{Q_{nm}^2}{2 C}
    \nonumber \\
    &+&e^{+t/\tau_C} \, \sum_{n,m} 
     \left[\frac{1}{2L} (\Phi_{nm} -\Phi_{ext} )^2
      -E_J \, \cos\left( 2\pi \frac{\Phi_{nm}}{\Phi_0} \right) \right.
  \nonumber \\
   &-&\frac{\lambda_x}{L} (\Phi_{nm} -\Phi_{ext} ) (\Phi_{n-1,m} -\Phi_{ext} )
  \nonumber \\
   &-&\frac{\lambda_y}{L} (\Phi_{nm} -\Phi_{ext} ) (\Phi_{n,m-1} -\Phi_{ext} ) 
    \left. \right] ,
\end{eqnarray}
where $E_J \equiv I_c\, \Phi_0 / 2\pi$ is the Josephson energy, $\tau_C=R\, C$,
and 
\begin{eqnarray}
  \label{8}
    Q_{nm} = e^{+t/\tau_C} \, C\, \frac{d \Phi_{nm}}{dt}
\end{eqnarray}
is the canonical variable conjugate to $\Phi_{nm}$, and represents the charge
accumulating across the capacitance of the JJ of each rf SQUID.
The Hamiltonian Eq. (\ref{7}) is a generalization in the 2D lossy case 
of that in Refs. \cite{Roscilde,Corato} used in the context of quantum 
computation with rf SQUID qubits.

\section{Single rf SQUID oscillator
}
The dynamic equation for a single rf SQUID is given from Eqs. (\ref{6})
for $\lambda_x = \lambda_y =0$ and $f_{nm} \rightarrow f$ 
\begin{eqnarray}
  \label{8.1}
   \frac{d^2 f}{d\tau^2} +\gamma \frac{d f}{d\tau} +f
   +\beta\, \sin( 2 \pi f ) = f_{ext} ,
\end{eqnarray}		
which has been studied extensivelly for more than two decades
both in the hysteretic ($\beta_L >1$) and the non-hysteretic 
($\beta_L <1$) regimes. 
The external driving $f_{ext}$ can be any time-dependent function,
which may also include a constant term. In the following, we assume
that the external flux is of the form 
\begin{eqnarray}
  \label{11}
     f_{ext} = f_{DC}  + f_{e0} \cos(\Omega \tau ) ,
\end{eqnarray}
where $f_{e0} = \Phi_{e0}/\Phi_0$ and $f_{DC} = \Phi_{DC}/\Phi_0$,
with $\Phi_{DC}$ being a constant (DC) flux resulting from
a time-independent and spatially uniform magnetic field.
The nonlinear dynamics of Eq. (\ref{8.1}) with $f_{ext}$ 
given by Eq. (\ref{11}) (with or without the DC term) is very rich, 
exhibiting bifurcations
and chaos in large portions of the parameter space
\cite{Soerensen,Ritala,Fesser}.
\begin{figure}[!t]
\includegraphics[angle=-0, width=.7\linewidth]{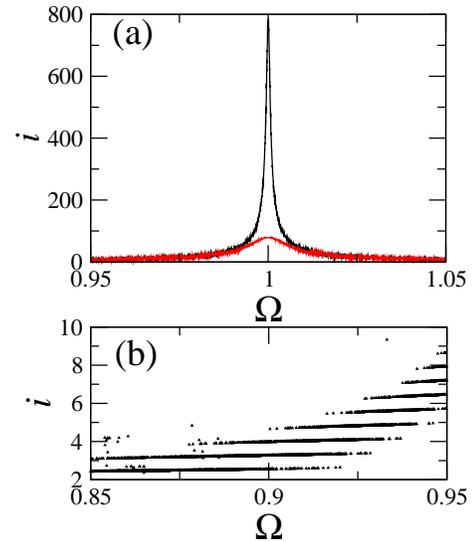}
\caption{
(a) The resonance curve of the induced (super)current $i$ as a function 
of the frequency of the applied 
rf field $\Omega$ for a single rf SQUID with 
$\beta=1.27$, $f_{DC}=0$, 
$f_{e0}=1.0$, and $\gamma=0.001$ (high peaked curve); 
$\gamma=0.01$ (low peaked curve).
(b) Enlargement of a small region from Fig. 2a where several fine steps 
on the resonance curve are clearly observable.
}
\end{figure}

The properties of rf SQUIDs in an alternating external field as a 
nonlinear resonant oscillator have been investigated experimentally both
for in hysteretic and the nonhysteretic (dispersive) regimes
\cite{Shnyrkov,Dmitrenko,Zeng}. The signal amplitude of the rf SQUID
as a function of the frequency of the applied rf field exhibits 
a strong resonance at a specific frequency at (or close to) $\omega_0$.
Although here we focus on the DDB generation in rf SQUID arrays,
we shall refer shortly to the peculiar resonance behaviour
of the single rf SQUID.
A typical resonance curve for a hysteretic rf SQUID is shown in Fig. 2a,
for two different values of the damping coefficient $\gamma$. 
Those symmetric, bell-shaped curves represent the flux amplitude of 
$f(\tau)$ as a function of the frequency $\omega$ of the applied field.
Apparently, those curves have a maximum at $\omega = \omega_0$, 
and they are similar to those observed in high$-T_c$ rf SQUIDs in 
an alternating field (see for example figure 2d in Ref.  \cite{Zeng}).  
Although these curves look smooth at first glance, they actually 
show multivalued behaviour, with jumps observed as fine steps as shown
in Fig. Fig. 2b.
For even lower rf power, however, we get very different resonance
curves, as shown in Fig. 3. Here, instead of the symmetric curves of 
Fig. 2a, which are characteristic of linear resonance 
(in case we forget the fine steps), 
we see a curve with a hysteretic loop,
which most closely resembles a typical nonlinear resonance curve. 
The hysteresis loops in Fig. 3 become smaller with increasing damping 
coefficient, as it could be expected.
Thus, for such low rf powers,
we can access the lowest energy states of the rf SQUID which 
actually are greatly affected  by the nonlinearity.  This behaviour
is peculiar to the rf SQUID, where nonlinear effects are stronger
for low applied power.

\begin{figure}[!t]
\includegraphics[angle=-0, width=.7\linewidth]{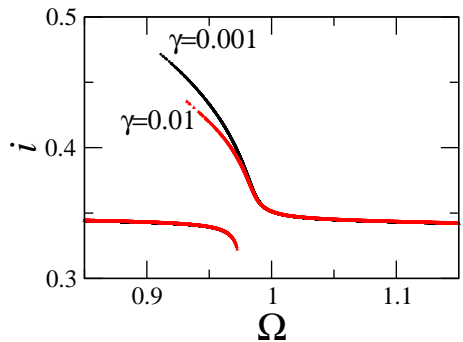}
\caption{
The resonance curve of the induced (super)current $i$ as a function 
of the frequency of the applied rf field $\Omega$ 
for a single rf SQUID with $\beta=1.27$, $f_{DC}=0$, $f_{e0}=0.5$, 
and $\gamma=0.001$ (high peaked curve); $\gamma=0.01$ (low peaked curve). 
For such low rf power, the lowest energy but highly nonlinear state
is accessed, which exhibits a typical nonlinear resonance curve with
a hysteretic loop. 
}
\end{figure}

\section{Linear dispersion.
}
The linear dispersion for small amplitude flux waves is obtained 
by the substitution of  
$f= A\, \exp[i (\kappa_x n + \kappa_y m - \Omega \tau)]$,
into the linearized Eq. (\ref{6}) without losses and external field 
($\gamma=0$, $f_{ext} =0$)
\begin{eqnarray}
  \label{9}
   \Omega = \sqrt{1 + \beta_L -2( \lambda_x \, \cos \kappa_x
                                    +\lambda_y \, \cos \kappa_y ) }  , 
\end{eqnarray}
where $\kappa_{x,y} = d_{x,y} \, k_{x,y}$ and $\Omega = \omega / \omega_0$.
The earlier equation describes the dispersion of a new kind of guided 
waves, the magneto-inductive (MI) waves, which are supported by
periodic, discrete arrays of magnetically coupled resonant elements \cite{Syms}.
Considering a 1D array, the corresponding dispersion 
(obtained by setting $\lambda_y=0$ and dropping the subscript $m$) 
has similar form with that
of electroinductive waves in chains of complementary metamaterial 
elements \cite{Beruete}. 
Moreover, in the limit of weak coupling ($\lambda_{x,y} \ll 1$), 
the dispersion (\ref{9}) has similar form with that
obtained for planar MI wave transducers, both in 
one \cite{Freire} and two \cite{Syms} dimensions.
Typical $\Omega(\kappa)$ curves are shown in Fig. 4a for three different
values of the coupling coefficient $\lambda=\lambda_x$.
The bandwidth $\Delta\Omega \equiv \Omega_{max} - \Omega_{min}$
decreases with decreasing $\lambda_x$ which
leads, for realistic values of $\lambda_x$ (between 0.05 and 0.1),
to a nearly flat band with $\Delta\Omega \simeq 2 \lambda \sqrt{1+\beta_L}$
(and relative bandwidth $\Delta\Omega /\Omega \simeq 2 \lambda$).
The corresponding phase and group velocities $v_{ph}$ and $v_g$, respectively,
for the red-dashed curve
of Fig. 4a are shown in Fig. 4b. 
(Notice that the actual value of $v_g$ has been multiplied by 250.)
Importantly, the group velocity $v_g$, which defines the 
direction of power flow, is in a direction opposite
to the phase velocity $v_{ph}$. 
Typical dispersion curves (i.e., contours of the frequency as a function of 
$\kappa_x$ and $\kappa_y$)
for both isotropic and anisotropic two-dimensional (2D) SQUID arrays
are shown in Figs. 5a and 5b, respectively.
In that case, $v_g$ is not, in general, in a direction opposite to $v_{ph}$.
\begin{figure}[!t]
\includegraphics[angle=-0, width=.7\linewidth]{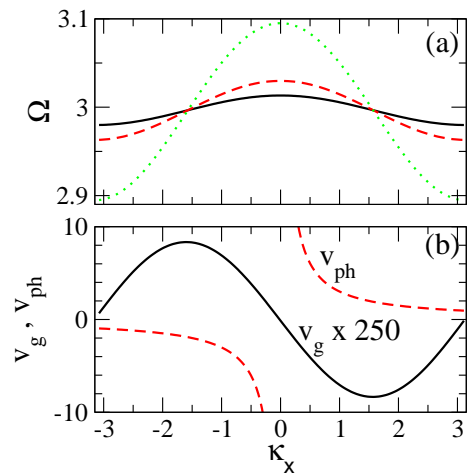}
\caption{
(a) Frequency band $\Omega$ as a function of $\kappa_x$ for a 1D rf SQUID array, 
for $\beta=1.27$, and $\lambda_x=-0.05$ (narrowest band, black solid curve),
$\lambda_x=-0.1$ (red dashed curve), 
$\lambda_x=-0.3$ (widest band, green dotted curve).
(b) Group velocity $v_g$ (black solid curve) 
and phase velocity $v_{ph}$ (red dotted curve), for a 1D rf SQUID array
with $\beta=1.27$ and $\lambda_x=-0.1$.
}
\end{figure}
\begin{figure}[!h]
\center{
\includegraphics[angle=-0, width=.4\linewidth]{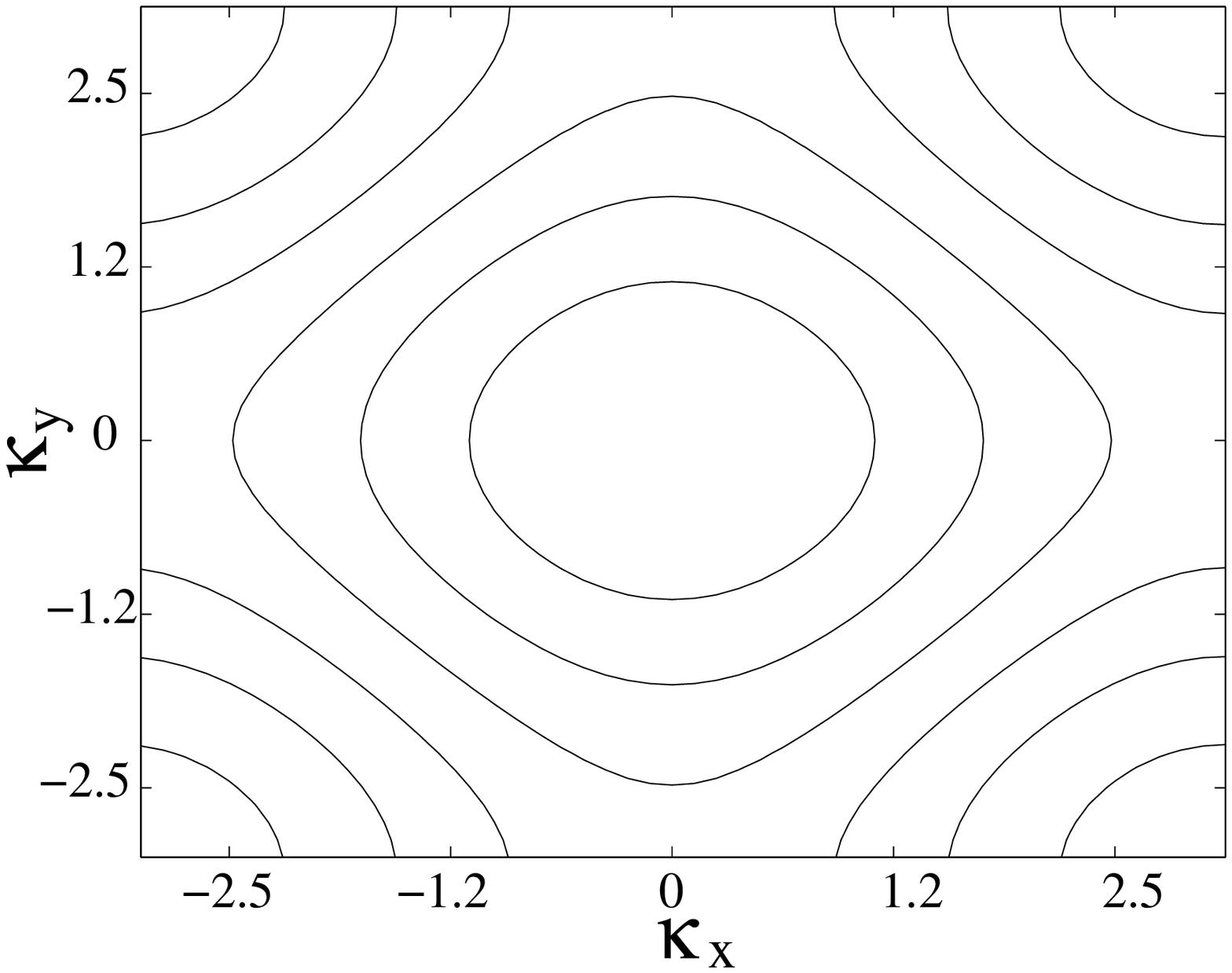}
\includegraphics[angle=-0, width=.4\linewidth]{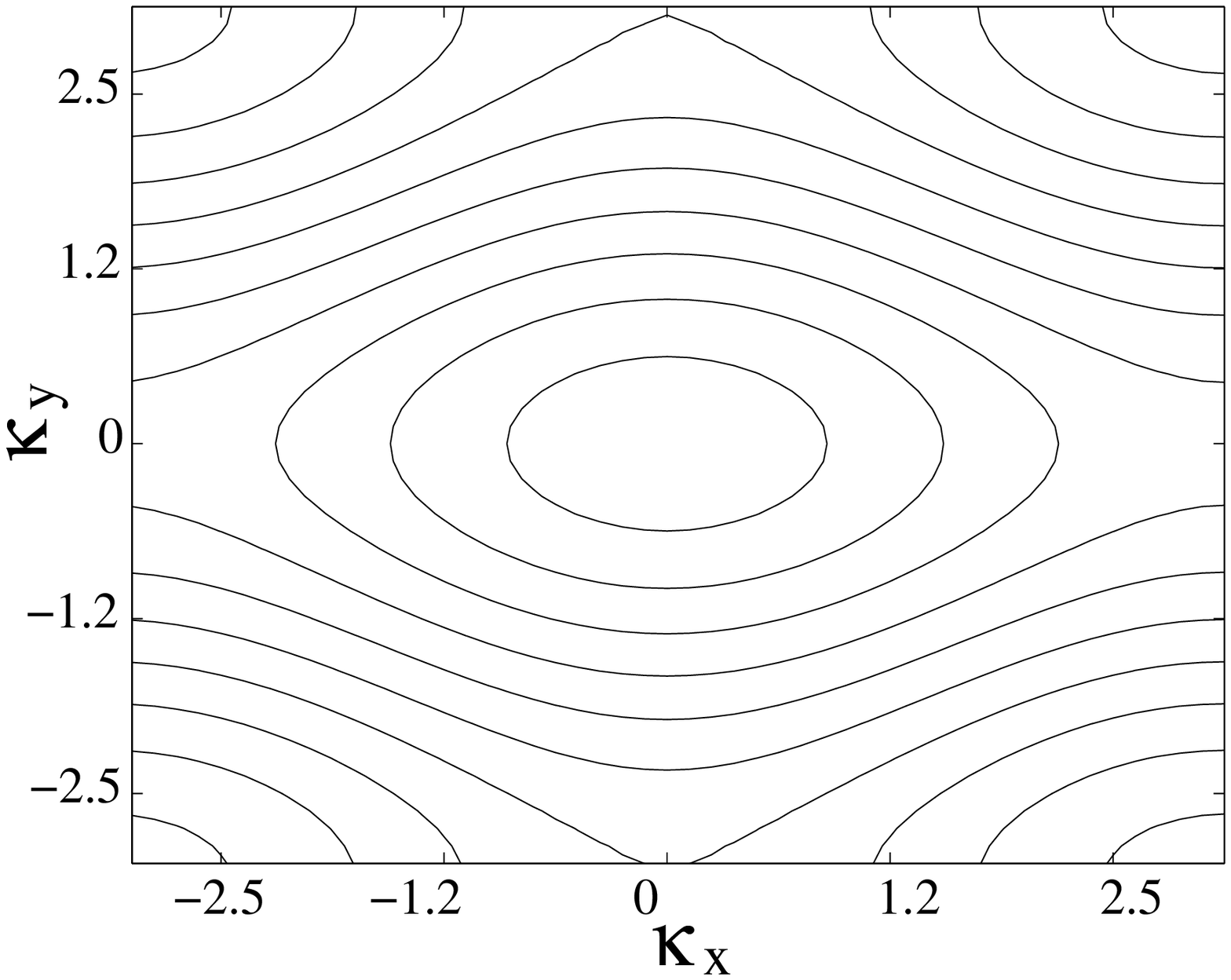}
}
\caption{
Contours of the linear dispersion $\Omega_{\vec\kappa}$ 
in the $\kappa_x - \kappa_y$ plane
for a two-dimensional rf SQUID array, with 
(i) $\lambda_x=-0.05$, $\lambda_y=-0.05$ (isotropic coupling);
(ii) $\lambda_x=-0.05$, $\lambda_y=-0.10$ (anisotropic coupling).
($\beta=1.27$).
}
\end{figure}

\section{Dissipative discrete breathers.
}
Consider first the simpler case of a 1D finite rf SQUID array,
consisting of $N$ identical units.
In order to generate DDBs we start by solving Eq. (\ref{6}) 
in the anti-continuous limit \cite{Marin},
i.e., for $\lambda_x \equiv \lambda=0$ when all SQUIDs are uncoupled.
Then, the 1D Eqs. (\ref{6}) reduce to Eq. (\ref{8.1}), 
the equation for a single damped and driven rf SQUID \cite{Likharev}. 
We identify two different amplitude coexisting and stable attractors 
of the single rf SQUID oscillator, with flux amplitudes $f_h$ and $f_\ell$
for the high and low amplitude attractor, respectively,
and corresponding voltages $v_h$ and $v_\ell$, respectively.
Subsequently, we fix the flux amplitude and the voltage of one of the rf SQUIDs 
(say the one at $n=n_b=N/2$) to  $f_h$ and $v_h$, respectively,
and all the others to $f_\ell$ and $v_\ell$, respectively.
Using this configuration (usually referred to as "trivial breather") 
as initial condition, 
we integrate the 1D Eqs. (\ref{6}) for a sufficiently small value of 
$\lambda=\delta\lambda$.
After integrating for a few hundred periods of the alternating driving field, 
the system has approached a stationary state.
Then, we again increase  $\lambda$ by $\delta\lambda$ 
and start to integrate again the 1D Eqs. (\ref{6}), using 
as initial condition the previously obtained stationary state.
After integrating for a few more hundreds driving periods,
the system has approached again a stationary state. Then,
we increase again $\lambda$ by $\delta\lambda$ and so on.
Using this algorithm, we can construct DDBs up to some maximum value of 
the coupling $\lambda$ \cite{Marin}.
For the integration of Eqs. (\ref{6}) we use a standard fourth-order
Runge-Kutta algorithm with fixed time-stepping $\Delta t$ 
(typically  $\Delta t =0.01$).
Since the DDBs presented here are highly localized,
the choice of boundary conditions to be imposed on Eqs. (\ref{6}) is not 
especially important.
Thus, we have chosen periodic boundary conditions throughout the study.
In the anti-continuous limit, all the SQUIDs are subjected to the same 
potential
\begin{figure}[t]
\includegraphics[angle=-0, width=.75\linewidth]{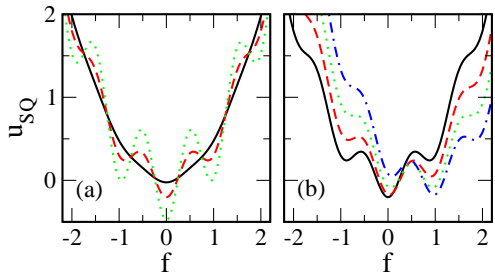}
\caption{
Potential function of a single rf SQUID $u_{SQ}$ as a function 
of its flux $f$, for
(a) $f_{ext}=0$, and $\beta=0.15$ (black solid curve),
$\beta=1.27$ (red dashed curve), $\beta=3$ (green dotted curve);
(b) $\beta=1.27$, and $f_{ext}=0$ (black solid curve),
$f_{ext}=0.25$ (red dashed curve), $f_{ext}=0.5$ (green dotted curve),
$f_{ext}=0.75$ (blue dashed-dotted curve).
}
\end{figure}
\begin{eqnarray}
  \label{10}
   u_{SQ} = \frac{1}{2} (f - f_{ext})^2 -\frac{\beta}{2\pi} \cos(2\pi f) .
\end{eqnarray}
Due to the form of $u_{SQ}$ shown in Fig. 6, it is a rather obvious task 
to construct a "trivial breather",
i.e., a DDB for $\lambda_x=\lambda_y=0$, when there are 
more than one local minimae. For example, when $\beta=1.27$ and $f_{DC} =0$
(read-dashed curve in Fig. 6a) 
one may choose the approximate values $f_\ell \simeq 0$ and $f_h \simeq 1$ 
(with $v_h =v_\ell \simeq 0$.  These values lead to stable states of the single 
rf SQUID equation which are localized into the left and the right local 
minimae of the potential, respectivelly.
Additionally, the choice of $f_{e0}$ should be such that both those states
will remain localize around the corresponding  local minimum.
By continuation of 
this trivial DDB for $\lambda_x , \lambda_y \neq 0$ one may obtain DDBs 
up to relatively high values of the coupling coefficients, 
whose maximum depends on the specific value of $f_{e0}$.
Such a DDB in a 1D rf SQUID array is shown in Fig. 7, 
where the spatio-temporal evolution of the induced currents $i_n$ ($n=1,2,3,...,N$)
are shown during one DDB period.
Both the background and the central DDB site are oscillating with 
the same frequency $\Omega_b =2\pi/T_b =\Omega$, 
i.e., a frequency equal to the driving frequency. 
We should also notice in Fig. 5 the non-sinusoidal time-dependence
of the oscillations.
When there are more than two local minimae in $u_{SQ}$ 
(e.g., for $\beta=3$, green-dotted curve in Fig. 6a)
we can construct more than one different DDBs, by combining any two of the 
different coexisting and stable states.  
These DDBs are extremely stable, since they are constructed
from stable and/or metastable (with very long life-time) localized states of $u_{SQ}$,
which can survive down to zero frequencies.
This is a characteristic example of a topological DDB.
\begin{figure}[!t]
\includegraphics[angle=-0, width=.8\linewidth]{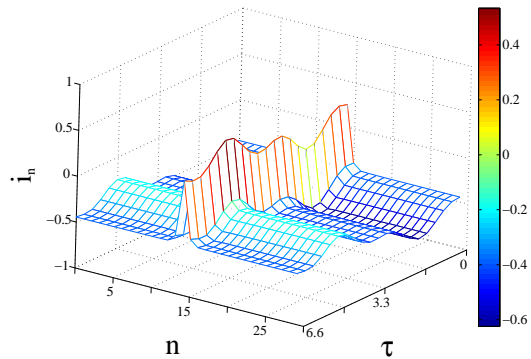}
\caption{
Time evolution of a discrete dissipative breather during one period,
for $f_{DC}=0.5$, $f_{e0}=0.2$,
$\beta=1.27$, $\alpha=0.001$, $\lambda=0.1$, and $T_b=6.6$.
Only part of the array ($N=30$) is shown for clarity. 
}
\end{figure}
\begin{figure}[!h]
\includegraphics[angle=-0, width=.8\linewidth]{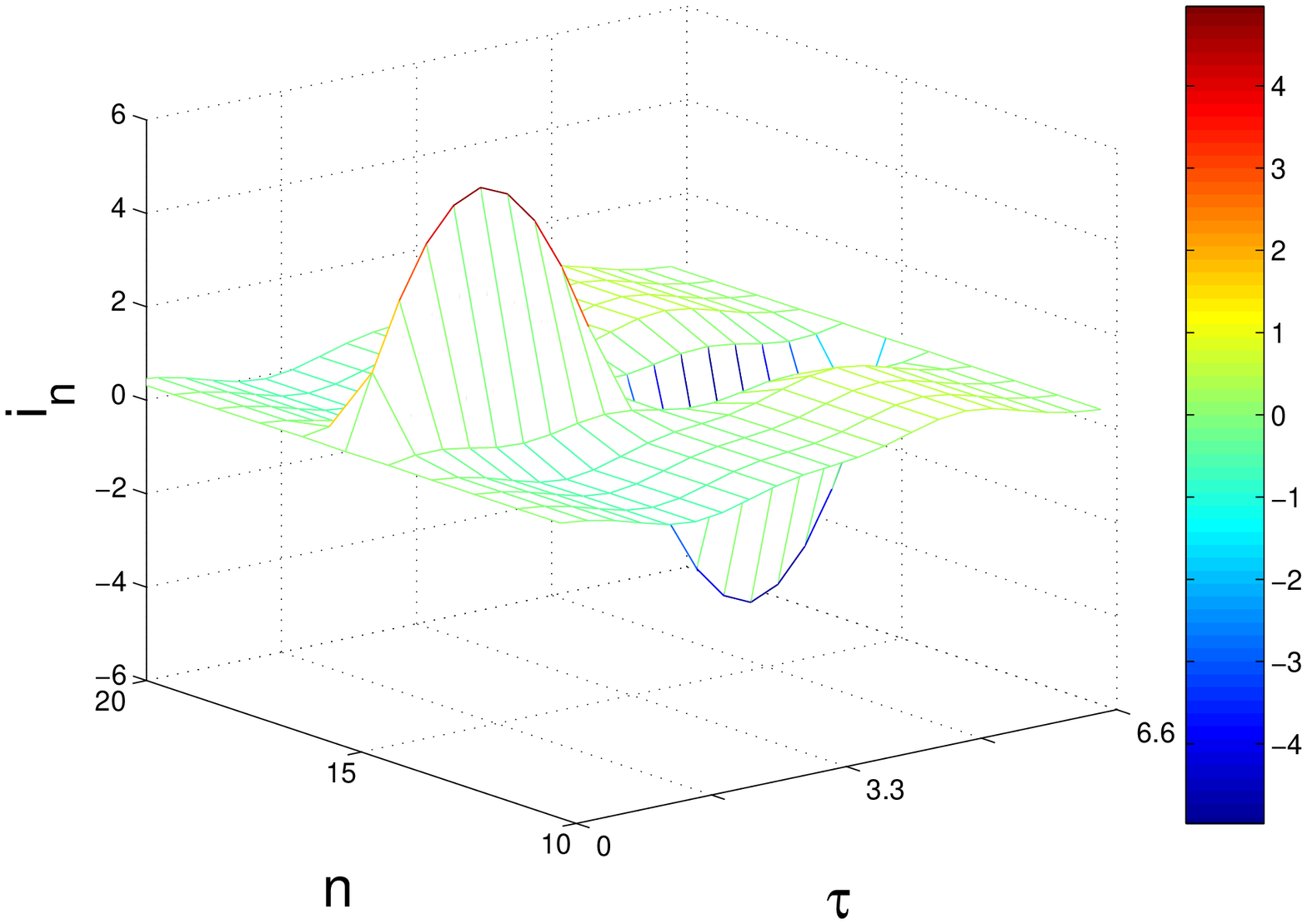}
\includegraphics[angle=-0, width=.8\linewidth]{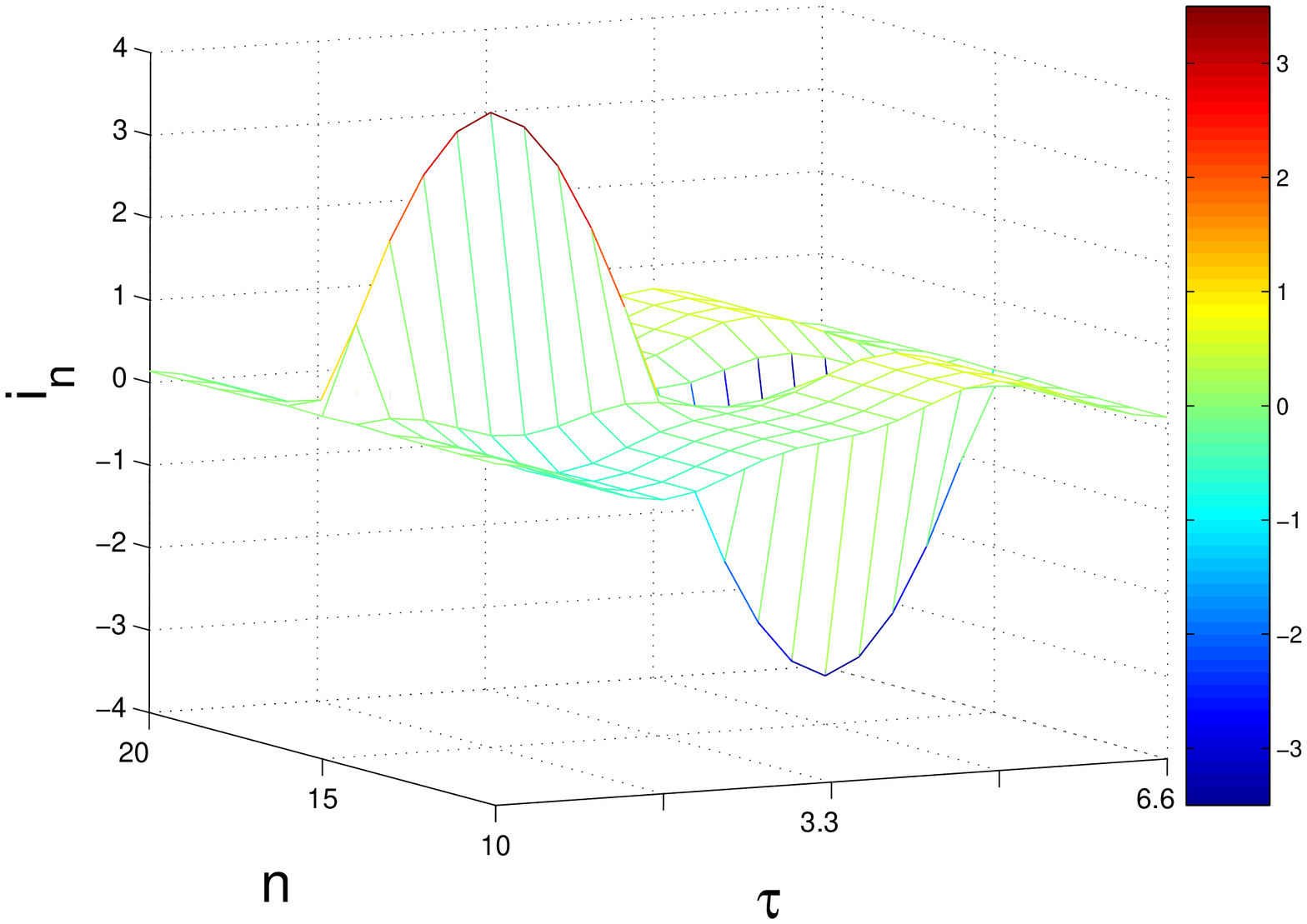}
\caption{
Time evolution of two different discrete dissipative breathers during 
one period,
for $f_{DC}=0$, $f_{e0}=0.6$,
$\beta=1.27$, $\alpha=0.001$, $\lambda=0.1$, and $T_b=6.6$.
Only part of the array ($N=30$) is shown for clarity. 
}
\end{figure}
However, one can also construct DDBs whose central site crosses 
the zero level while oscillating. This requires the use of
high amplitude states of the single rf SQUID oscillator,
which may turn the SQUID into the normal (i.e., not superconducting)
state. At that state, a rather large voltage difference appears
along the JJ of that SQUID.
Two typical examples of such DDBs, which may coexist, are shown in Fig. 8.
Both the background and the central DDB site oscillate with the same frequency
but different amplitudes
(low and high amplitude current oscillation, respectivelly).
The frequency of the oscillations, and thus the DDB frequency $\Omega_b$,
is again equal to the driving frequency $\Omega$ ($\Omega_b=\Omega$). 
However, there is a difference between the phases of the oscillation
between the background and the central DDB site which is almost $\pi$, 
and that has profound consequences in the local magnetic properties
of the array (see below).
Although here we present only one-site, bright dissipative DBs,
we can contruct, by choosing appropriate initial conditions,
many different types of DDBs.
The linear stability of DDBs is addressed
through the eigenvalues of the Floquet matrix (Floquet multipliers).
A DDB is linearly stable when all its Floquet multipliers 
$m_i, ~i=1,...,2N$ lie on a circle of radius $R_e = \exp(-\alpha T_b/2)$ 
in the complex plane \cite{Marin1}.
The DDBs shown in Figs. 7 and 8 (as well as those shown below),
are all linearly stable. 
The calculated eigenvalues for the DDBs presented in Figs. and 8a and 8b 
are shown in the complex plane in Fig. 9a and 9b, respectivelly.
Moreover, those DDBs were let to evolve for large
time intervals (i.e., more than $10^5~T_b$) without any observable change 
in their shapes.\vspace{1.5cm}
\begin{figure}[!h]
\includegraphics[angle=-0, width=.7\linewidth]{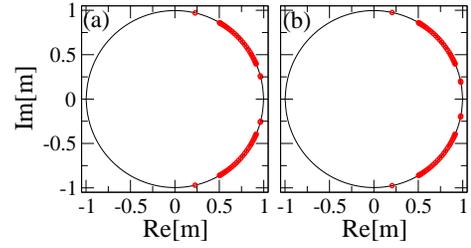}
\caption{
Floquet spectra for the one-site bright dissipative breathers
shown in (a) the upper panel of Fig. 6;
and (b) the lower panel of Fig. 6.
All eigenvalues lie on a circle of radius 
$R_e = \exp(-\alpha T_b/2) \simeq 0.996705$ in the complex plane.
}
\end{figure}

We can also construct DDBs with periods which are
multiple of the that of the external driver (subharmonic DDBs), 
for relatively weak coupling.
Such a period-3 DDB, which is linearly stable, is shown in Fig. 10,
while the Floquet spectrum of its eigenvalues is shown in Fig. 11a.
In order to check directly its stability, this DDB was let to evolve for 
more than $5\times 10^5~T_b$, without any observable change of its profile.  
We conclude, thus, that this period-3
DDB is stable, or at least that it is very long-lived.
In Fig. 11b we show the Poincar\'e diagram 
(i.e., a diagram of $f_n$ vs. $v_n=d f_n/d\tau$ at the end of each period
of the driver),
for the central DDB site ($n=n_b=N/2$), as well as the site at $n=7$ 
which is located in the background.
Clearly, the trajectory of the central DDB site crosses the Poincar\'e surface at 
three points (red circles), while that of the site in the background at 
one point (black square).
\begin{figure}[!t]
\includegraphics[angle=-0, width=.8\linewidth]{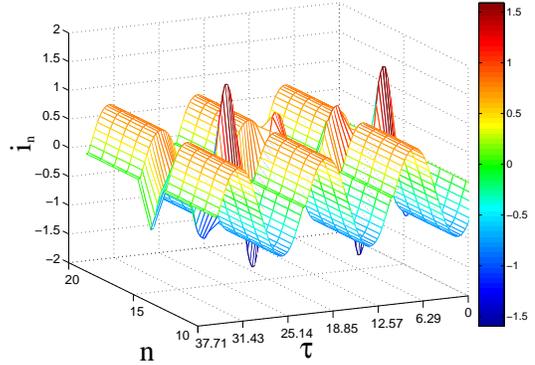}
\caption{
Time evolution of a discrete, dissipative, period-3 breather during three
driver periods,
for $f_{DC}=0$, $f_{e0}=1.2$,
$\beta=1.27$, $\alpha=0.001$, $\lambda=0.0225$, and $T_b=12.57$.
Only part of the simulated array ($N=30$) is shown for clarity. 
}
\end{figure}

Most of the methodology and techniques for DB construction
has been developed for the 1D case. However, a rigorous proof of the 
existence of DBs in higher-dimensional nonlinear lattices was given in 
\cite{Mackay}, and several numerical studies of DBs in 2D nonlinear
lattices have been published \cite{Flach2,Mazo,Burlakov,Kevrekidis}.
Since rf SQUID arrays are fabricated in planar (2D) technology,
it is necessary to extend the study of MI-DDBs in
these systems in two dimensions.
We have seen that
DDBs are not destroyed by increasing the dimensionality from one to two.
Consider a 2D $N\times N$ rf SQUID array consisting of identical units.
Following the same procedure that we used to construct one-dimensional DDBs,
we start from the anti-continuous limit by solving the single rf SQUID
equation with losses and a driving term of the form of Eq. (\ref{11}). 
We identify two different
coexisting and stable attractors of that oscillator with flux amplitudes 
$f_h$ and $f_\ell$ and corresponding voltages $v_h$ and $v_\ell$, respectivelly.
Then, in order to construct a trivial breather, 
we fix the flux amplitude and voltage of one of the rf SQUIDs
(say the one at $(n,m)=(n_b,n_b)=(N/2,N/2)$) 
to  $f_h$ and $v_h$, respectively,  
and all the others to $f_\ell$ and $v_\ell$, respectively.
Then we integrate the 2D system of Eqs. (\ref{6}) while increasing 
simultaneously the coupling coefficients $\lambda_x$ and $\lambda_y$ 
in small steps, as it was described earlier.
Using this algorithm we have constructed several DDBs for an isotropic rf 
SQUID array
($\lambda_x = \lambda_y=\lambda$), up to some maximum $\lambda$.
A snapshot of such a typical 2D isotropic DDB profile 
(at maximum amplitude of the central site)
is shown in Fig. 12, for the same parameters used to construct the 
one-dimensional DDB of the top panel of Fig. 8.
Note that the coupling coefficients $\lambda_x$ and $\lambda_y$ may differ 
in magnitude leading to anisotropic rf SQUID arrays.
We have also constructed DDBs  in anisotropic 2D arrays
where $\lambda_x \neq \lambda_y$ (not presented here), for a wide range of  
the anisotropy parameter $\lambda_y / \lambda_x$.\vspace{1.2cm}
\begin{figure}[!t]
\includegraphics[angle=-0, width=.65\linewidth]{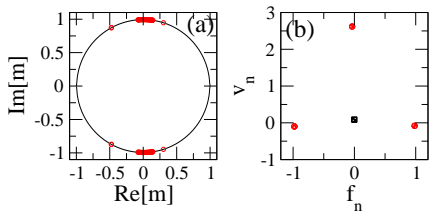}
\caption{
(a) Floquet spectra for the one-site bright, dissipative, period-3 breather
shown in fig. 8; all eigenvalues are in a circle of radius 
$R_e = \exp(-\alpha T_b/2) \simeq 0.993735$.
(b) Poincar\'e surface of section for the central DB site at $n=n_b$
(red circles), and the site at $n=7$ located in the background (black square),
of the  period-3 breather shown in fig. 8.
}
\end{figure}

\section{Magnetic response.
}
It is apparent from Fig. 8 that the low and high amplitude current oscillators
have different phases with respect to the applied magnetic field.
Consequently, their magnetic response in that field is expected to be different. 
To see that, we cast the normalized Eq. (\ref{1}) in the form 
\begin{equation}
 \label{12}
   \beta \, i_{nm} = f_{nm}^{loc} - f_{ext}^{eff} ,
\end{equation}
where 
\begin{eqnarray}
\label{13}
  f_{nm}^{loc} &=& f_{nm} 
  \nonumber \\
      &-&\lambda_x ( f_{n-1,m} + f_{n+1,m} )
      -\lambda_y ( f_{n,m-1} + f_{n,m+1} ) , \nonumber \\
      \\
  \label{14}
  f_{ext}^{eff} &=& [1 -2 (\lambda_x + \lambda_y)] f_{ext} .
\end{eqnarray}
\begin{figure}[!t]
\includegraphics[angle=-0, width=.8\linewidth]{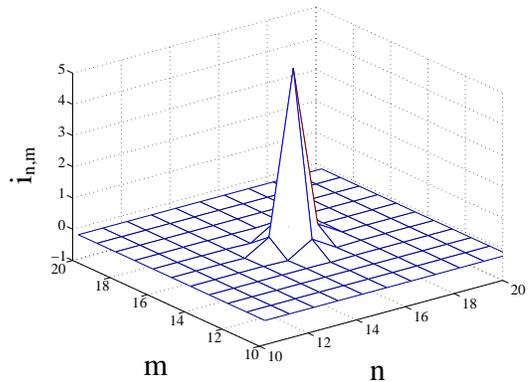}
\caption{
A snapshot of a dissipative discrete breather (at maximum amplitude
of the central site) for the parameters of fig. 6.
Only part of the simulated array ($30\times 30$) is shown for clarity.
}
\end{figure}
After division by the area of the unit cell $d^2$ of the 2D array,
the terms $f_{ext}^{eff}$, $f_{nm}^{loc}$, and
$\beta \, i_{nm}$ in (\ref{12}) can be interpreted as the effective 
external (driving) field,
the local magnetic induction at cell ($n,m$), and the magnetic response 
(magnetization) at cell ($n,m$), respectivelly.
Consider the DB shown in the top panel of Fig. 10.
The temporal evolution (during one period) of $\beta \, i_{nm}$, $f_{nm}^{loc}$,
and $f_{ext}^{eff}$, which are directly proportional to 
its magnetic response, the local magnetic induction, and the external magnetic
field, respectivelly, are shown in Fig. 13 for two different cells of the array;
the central (breather) cell at $(n,m)=(n_b,n_b)$, 
and the cell at $(n,m)=(7,7)$ (Fig. 13a and Fig. 13b, respectively).
The latter is chosen to lie in the oscillating background, relatively far from the
central DDB site and the ends of the array.
We observe significant differences in the magnetization 
(red-solid curves) in those two cells;
in the cell containing the high current amplitude oscillator 
(i.e., the central DDB site)
the magnetization is in phase with the applied field,
while in the other cell the magnetization is in anti-phase with that. 
Thus, in the present case, the DDB provides a diamagnetic response in a 
strongly paramagnetic background. 
In this sence, a DDB may alter locally the character 
of the magnetic response (paramagnetic/diamagnetic) of a SQUID array in an 
alternating magnetic field. 
In some cases, the magnitude of the magnetization of the DDB 
may exceed that of the applied field, leading to extreme diamagnetic or negative
magnetic response.

\begin{figure}[!t]
\includegraphics[angle=-0, width=.7\linewidth]{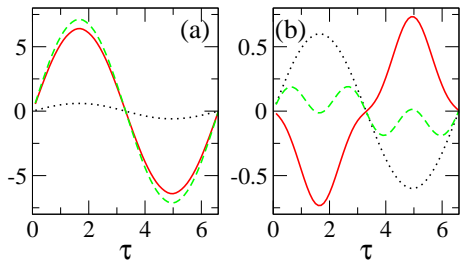}
\caption{
Temporal evolution of the response $\beta\, i_n$ (red-solid curve), 
the local flux $f_n^{loc}$ (green-dashed curve),
and the external flux $f_{ext}$ (black-dotted curve) 
during one period $T_b$, for
(a) the central site of the dissipative discrete breather shown in the top panel 
of Fig. 6 ($n=n_b=N/2$);
(b) the site with $n=7$ (which is located in the background) of the dissipative 
discrete
breather shown in the top panel of Fig. 6 ($n=7$).
}
\end{figure}
\section{Conclusions.
}
In conclusion, we have shown using standard numerical methods that both 1D 
and 2D periodic rf SQUID arrays 
in an alternating external flux support several types of linearly stable DDBs.
Those arrays belong to the class of MI systems, since the individual rf SQUIDs
are weakly coupled through magnetic interactions.
Similar MI DDBs were found to exist also in arrays of split-ring 
resonators \cite{Lazarides,Eleftheriou}, 
which constitute the most common elements for the construction of MMs.
We speculate that DDBs generically exist in discrete and nonlinear MI systems,
for rather wide parameter ranges, and they are linearly stable for weak coupling
between their units.
For the rf SQUID array we have also obtained different DDB excitations which may 
co-exist as well as multiperiodic DDBs, which are linearly stable. 
The latter are obtained only for relatively weak coupling between SQUIDs.
Moreover, DDBs may alter locally the magnetization (magnetic response)
of an rf SQUID array in an alternating magnetic field.
The increasing of dimensionality does not, in general, destroy the DDB solutions. 
Thus, it seems possible to exploit dissipative 
multibreathers in order to create strongly paramagnetic "islands" in a 2D SQUID
array, surrounded by a diamagnetic (or even extreme diamagnetic) background.  
The co-existence of several linearly stable DDB is a result of the rich
nonlinear dynamics of single SQUIDs, which allows for multistability
even for frequencies far from resonance. The weak coupling modifies only 
slightly the amplitude of oscillation of those states in each SQUID in the 
array.
Thus, it is also possible to get a multiplicity of uniform solutions in 
a wide range of frequencies,
which provide different magnetic responses (paramagnetic or diamagnetic).

\section*{Acknowledgements.
}
We acknowledge  support from the grant "Pythagoras II" (KA. 2102/TDY 25)
of the Greek Ministry of Education and the European Union.


\begin{references}

\bibitem{Flach}
  S. Flach and C. R. Willis.
  Phys. Rep. {\bf 295}, 181 (1998).
  
\bibitem{Campbell}
  D. K. Campbell, S. Flach, and Y. S. Kivshar.
  Physics Today 43, January (2004).

\bibitem{Peyrard}
  M. Peyrard.
  Physica D {\bf 119}, 184 (1998).

\bibitem{Rasmussen}
  K. {\O}. Rasmussen, S. Aubry, A. R. Bishop, and G. P. Tsironis.
  Eur. Phys. J. B {\bf 15}, 169 (2000).

\bibitem{Rasmussen1}
  K. {\O}. Rasmussen, D. Cai, A. R. Bishop, and N. Gr{\o}nbech-Jensen.
  Europhys. Lett. {\bf 47}, 421 (1999).

\bibitem{Hennig}
  D. Hennig, L. Schimansky-Geier, and P. H\"anggi.
 Europhys. Lett. {\bf 78}, 20002 (2007).

\bibitem{Hennig1}
  D. Hennig, S. Fugmann, L. Schimansky-Geier, and P. H\"anggi.
  Phys. Rev. E {\bf 76 (4)}, 041110 (2007).

\bibitem{Sievers}
  A. J. Sievers and S. Takeno.
  Phys. Rev. Lett. {\bf 61}, 970 (1988).

\bibitem{Mackay}
  R. S. MacKay and S. Aubry.
 Nonlinearity {\bf 7}, 1623 (1994).

\bibitem{Aubry}
  S. Aubry.
  Physica D {\bf 103}, 201 (1997).

\bibitem{Marin}
  J. L. Mar\'in and S. Aubry.
  Nonlinearity {\bf 9}, 1501 (1996).

\bibitem{Marin1}
  J. L. Mar\'in, F. Falo, P. J. Mart\'inez, and L. M. Flor\'ia.
  Phys. Rev. E {\bf 63}, 066603 (2001).
  
\bibitem{Zueco}
  D. Zueco, P. J. Mart\'{i}nez, L. M. Flor\'{i}a, and F. Falo.
 Phys. Rev. E {\bf 71}, 036613 (2005).

\bibitem{Bergamin}
  J. M. Bergamin and T. Bountis.
  Prog. Theor. Phys. Suppl. {\bf 150}, 330 (2003).

\bibitem{Panagopoulos}
  P. Panagopoulos, T. Bountis, and C. Skokos.
  J. Vib. Acoust. {\bf 126}, 520 (2004).

\bibitem{Tsironis}
  G. P. Tsironis.
  J. Phys. A:Math. Gen. {\bf 35}, 951 (2002).

\bibitem{Swanson}
  B. I. Swanson, J. A. Brozik, S. P. Love, G. F. Strouse, 
  A. P. Shreve,
  A. R. Bishop, W.-Z. Wang, and M. I. Salkola.
  Phys. Rev. Lett. {\bf 82}, 3288 (1999).

\bibitem{Schwarz}
  U. T. Schwarz, L. Q. English, and A. J. Sievers.
  Phys. Rev. Lett. {\bf 83}, 223 (1999).

\bibitem{Trias}
  E. Tr\'ias, J. J. Mazo, and T. P. Orlando.
  Phys. Rev. Lett. {\bf 84}, 741 (2000).

\bibitem{Sato}
  M. Sato, B. E. Hubbard, A. J. Sievers, B. Ilic, 
  D. A. Czaplewski, and H. G. Graighead.
  Phys. Rev. Lett. {\bf 90}, 044102 (2003).

\bibitem{Eisenberg}
  H. S. Eisenberg, Y. Silberberg, R. Morandotti, A. R. Boyd, 
  and J. S. Aitchison.
  Phys. Rev. Lett. {\bf 81}, 3383 (1998).

\bibitem{Russell}
  F. M. Russell, and J. C. Eilbeck.
  Europhys. Lett. {\bf 78}, 10004 (2007).

\bibitem{Edler}
  J. Edler, R. Pfister, V. Pouthier, C. Falvo, and P. Hamm.
  Phys. Rev. Lett. {\bf 93}, 106405 (2004).

\bibitem{Martinez1}
  P. J. Mart\'inez, L. M. Flor\'ia, F. Falo, and J. J. Mazo.
  Europhys. Lett. {\bf 45}, 444 (1999).

\bibitem{Maniadis}
  P. Maniadis and T. Bountis.
  Phys. Rev. E {\bf 73}, 046211 (2006).

\bibitem{Lazarides}
  N. Lazarides, M. Eleftheriou, and G. P. Tsironis.
  Phys. Rev. Lett. {\bf 97}, 157406 (2006).
    
\bibitem{Eleftheriou}
  M. Eleftheriou, N. Lazarides, and G. P. Tsironis.
  "Magnetoinductive breathers in magnetic metamaterials",
  submitted to Phys. Rev. E, October 2007. 
  e-print: arXiv:0709.3567 [cond-mat.mtrl-sci] 22 Sep 2007.

\bibitem{Yen}
  T. J. Yen, W. J. Padilla, N. Fang, D. C. Vier, D. R. Smith,
  J. B. Pendry, D. N. Basov, and X. Zhang.
  Science {\bf 303}, 1494 (2004).

\bibitem{Podolskiy}
  V. A. Podolskiy, A. K. Sarychev and V. M. Shalaev.
  Opt. Express {\bf 11}, 735 (2003).

\bibitem{Soukoulis}
   C. M. Soukoulis, S. Linden, M. Wegener.
   Science {\bf 315}, 47 (2007).

\bibitem{Lazarides1}
  N. Lazarides, and G. P. Tsironis,
  Appl. Phys. Lett. {\bf 16}, 163501 (2007).
  
\bibitem{Likharev}
  K. K. Likharev.
  {\em Dynamics of Josephson Junctions and Circuits.}
  (Gordon and Breach, Philadelphia, 1986).

\bibitem{Barone}
  A. Barone and G. Pattern\'o.
  {\em Physics and Applications of the Josephson Effect.},
  (Wiley, New York, 1982).

\bibitem{Josephson}
  B. Josephson.
  Phys. Lett. A {\bf 1}, 251 (1962).

\bibitem{Chen1}
  D.-X. Chen, J. J. Moreno, A. Hernando and A. Sanchez.
  Europhys. Lett. {\bf 41}, 413 (1998).
  
\bibitem{Kourakis}
  I. Kourakis, N. Lazarides, and G. P. Tsironis.
  Phys. Rev. E {\bf 75}, 067601 (2007).

\bibitem{Roscilde}
  T. Roscilde, V. Corato, B. Ruggiero, and P. Silvestrini.
  Phys. Lett. A {\bf 345}, 224 (2005).
  
\bibitem{Corato}
  V. Corato, T. Roscilde, B. Ruggiero, C. Granata, and
  P. Silvestrini.
  J. Phys: Conf. Series {\bf 43}, 1401 (2006).

\bibitem{Soerensen}
  M. P. S{\o}rensen, M. Bartuccelli, P. L. Christiansen,
  and A. R. Bishop.
  Phys. Lett. A {\bf 109}, 347 (1985).
 
\bibitem{Ritala}
  R. K. Ritala and M. M. Salomaa.
  Phys. Rev. B  {\bf 29}, 6143 (1984). 

\bibitem{Fesser}
  K. Fesser, A. R. Bishop and P. Kumar.
  Appl. Phys. Lett. {\bf 43}, 123 (1983).

\bibitem{Shnyrkov}
  V. I. Shnyrkov, V. A. Khlus and G. M. Choi.
  J. Low Temp. Phys. {\bf 39}, 447 (1980).

\bibitem{Dmitrenko}
  I. M. Dmitrenko, G. M. Choi, V. I. Shnyrkov and V. V. Kartsovnik.
  J. Low Temp. Phys. {\bf 49}, 417 (1982).

\bibitem{Zeng}
  X. H. Zeng, Y. Zhang, B. Chesca, K. Barthel, Ya. S. Greenberg
  and A. I. Braginski.
  J. Appl. Phys. {\bf 88}, 6781 (2000).

\bibitem{Syms}
  R. R. A. Syms, E. Shamonina, and L. Solymar.
  Eur. Phys. J. B {\bf 46}, 301 (2005).

\bibitem{Beruete}
  M. Beruete, F. Falcone,  M. J. Freire, R. Marqu\'es, and J. D. Baena.
  Appl. Phys. Lett. {\bf 88}, 083503 (2006).

\bibitem{Freire}
  M. J. Freire, R. Marqu\'es, F. Medina, M. A. G. Laso,
  and F. Mart\'in.
  Appl. Phys. Lett. {\bf 85}, 4439 (2004).

\bibitem{Flach2}
  S. Flach, K. Kladko, and S. Takeno.
  Phys. Rev. Lett. {\bf 79}, 4838 (1997).

\bibitem{Mazo}
   J. J. Mazo.
  Phys. Rev. Lett. {\bf 89}, 234101 (2002).

\bibitem{Burlakov}
  V. M. Burlakov, S. A. Kiselev, and V. N. Pyrkov.
  Phys. Rev. B {\bf 42}, 4921 (1990).

\bibitem{Kevrekidis}
  P. G. Kevrekidis, K. {\O}. Rasmussen, and A. R. Bishop.
  Phys. Rev. E {\bf 61}, 2006 (2000).

\end{references}
\end{document}